\documentclass[12pt,preprint]{aastex}
\usepackage{natbib}
\begin{document}
\slugcomment{ApJ in press; submitted 2011.09.26}
\title{A Hot Gap Around Jupiter's Orbit in the Solar Nebula}

\author{N. J. Turner, M. Choukroun, J. Castillo-Rogez and G. Bryden}
\affil{Jet Propulsion Laboratory, California Institute of Technology,
  Pasadena, California 91109, USA; neal.turner@jpl.nasa.gov}

\begin{abstract}
  The Sun was an order of magnitude more luminous during the first few
  hundred thousand years of its existence, due in part to the
  gravitational energy released by material accreting from the Solar
  nebula.  If Jupiter was already near its present mass, the planet's
  tides opened an optically-thin gap in the nebula.  We show using
  Monte Carlo radiative transfer calculations that sunlight absorbed
  by the nebula and re-radiated into the gap raised temperatures well
  above the sublimation threshold for water ice, with potentially
  drastic consequences for the icy bodies in Jupiter's feeding zone.
  Bodies up to a meter in size were vaporized within a single orbit if
  the planet was near its present location during this early epoch.
  Dust particles lost their ice mantles, and planetesimals were
  partially to fully devolatilized, depending on their size.
  Scenarios in which Jupiter formed promptly, such as those involving
  a gravitational instability of the massive early nebula, must cope
  with the high temperatures.  Enriching Jupiter in the noble gases
  through delivery trapped in clathrate hydrates will be more
  difficult, but might be achieved by either forming the planet much
  further from the star, or capturing planetesimals at later epochs.
  The hot gap resulting from an early origin for Jupiter also would
  affect the surface compositions of any primordial Trojan asteroids.
\end{abstract}


\section{INTRODUCTION}

The Sun was more luminous when very young \citep{1961PASJ...13..450H,
  1994ApJS...90..467D, 2000A&A...358..593S} and it is natural to ask
about the consequences for planet formation.  Sunlight was the main
heat source for the outer parts of the protosolar nebula
\citep{2007prpl.conf..555D}, so the Solar luminosity surely affected
the distribution of ices.  Water ice was the most abundant solid in
locations where the temperatures were below its sublimation threshold
\citep{1994ApJ...421..615P, 2009Icar..200..672D} and played a key role
in the assembly of the outer planets, their moons, the Kuiper belt
objects and the comets, judging from the bodies' compositions.  Water
ice is invoked as a major building block for the solid cores of the
gas giant planets, whether the cores were built before
\citep{1996Icar..124...62P} or after \citep{2009ApJ...697.1256H} the
gaseous envelopes.  Water ice also is used to explain Jupiter's
atmospheric enrichment in the noble gases argon, krypton and xenon
relative to the Sun \citep{2003NewAR..47....1Y}.  Volatile species
including the noble gases can be trapped in amorphous ice forming at
temperatures below about 30~K \citep{1999Natur.402..269O} and in
clathrate hydrates at somewhat higher temperatures
\citep{1985ApJS...58..493L, 2004P&SS...52..623H}.  Extra motivation
for examining the delivery of ices to Jupiter is the prospect of
better composition and internal structure measurements after the {\sc
  Juno} spacecraft reaches the planet in 2016
\citep{2010IAUS..269...92B}.

In this paper we focus on the consequences of the Sun's higher early
luminosity for material close to Jupiter.  Ideas about Jupiter's
origins fall into two main groups.  In core-nucleated accretion
models, a solid core of ice and rock accumulates over one to several
million years before a significant gaseous envelope is accreted
\citep{1996Icar..124...62P, 2005Icar..179..415H, 2010Icar..209..616M}.
In gravitational instability models, the Solar nebula at some time
during its first few hundred thousand years is sufficiently massive to
fragment locally under its own self-gravity.  Jupiter forms within a
thousand years from one of the fragments \citep{1997Sci...276.1836B,
  2007prpl.conf..607D}.  The two groups of models thus differ
drastically in the epoch when the planet approaches its final mass.

When applied to the exoplanet population, core-nucleated accretion
more easily explains the formation of the gas giants with massive
cores \citep{2005ApJ...633..465S}, while gravitational instability can
account for the formation of high-mass exoplanets found far from their
host stars \citep{2009ApJ...707...79D, 2010ApJ...716L.176C}.

Once formed, Jupiter is sufficiently massive that its tides open a gap
in the Solar nebula \citep{1986ApJ...307..395L, 1993prpl.conf..749L}.
The reduced surface density in the gap may allow sunlight to penetrate
to the equatorial plane, either through scattering, or absorption in
the nebula and re-radiation.  Only in the gravitational instability
scenarios does Jupiter become massive enough to open such a gap in the
nebula while the Sun is still much more luminous than today.  Our goal
is to evaluate the impact of the high luminosity on the planet and its
surroundings.

Below, we compute the temperatures reached in the nebula and gap under
thermal equilibrium with the radiation from the protosun.  Similar
radiative transfer calculations by \cite{2006ApJ...637L.125V} revealed
that owing to the protostellar disk's concave or flared surface, the
outer edge of the gap intercepts an excess of starlight, and appears
as a bright ring in synthetic scattered light images.
\cite{2006ApJ...637L.125V} presented results with a fixed power-law
radial variation in the density scale height, and stated that the
outcome was similar under hydrostatic equilibrium.  Their central sun
had about $1.4 L_\odot$, corresponding to a T~Tauri star of age
1--3~Myr.  In contrast, our models all are in hydrostatic balance, and
we consider a protostar aged less than 0.3~Myr with ten times the
Solar luminosity, appropriate for the early epoch of planet formation
in the gravitational instability picture.  We find that the
more-luminous sun heats the gap enough to destroy many volatiles in
the vicinity of the planet.

The paper is laid out as follows.  The models for the protosun and
nebula are described in sections~\ref{sec:sun} and \ref{sec:nebula},
the radiative transfer approach in section~\ref{sec:rt} and the
iterative method used to obtain joint radiative and hydrostatic
equilibrium in section~\ref{sec:hse}.  The equilibrium results are
presented in section~\ref{sec:results}, the location and extent of the
heating and the effects on ices discussed in
section~\ref{sec:discussion}, and the implications for gas giant
planet formation summarized in section~\ref{sec:conclusions}.

\section{METHODS}

\subsection{The Sun
  \label{sec:sun}}

\begin{figure}[tb!]
  \epsscale{0.8} \plotone{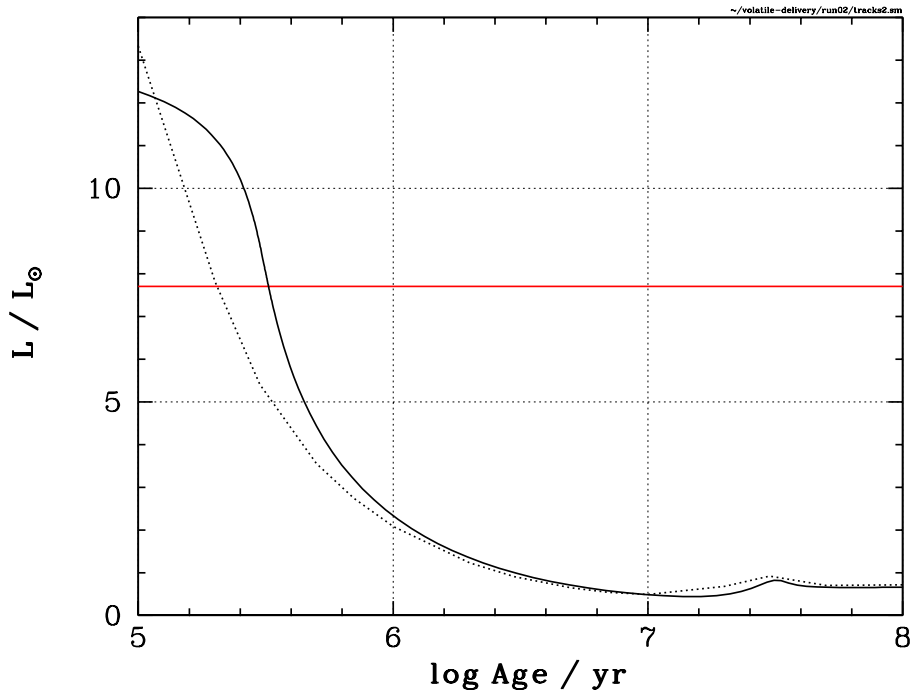} \figcaption{\sf Luminosity vs.\ time
    in two models for the evolution of a Solar-mass protostar.  The
    dotted curve is from \cite{1994ApJS...90..467D}, the solid curve
    from \cite{2000A&A...358..593S}.  Both models include nuclear
    fusion and the heat of contraction, but fix the stellar mass and
    neglect the energy released by ongoing accretion.  A horizontal
    red line shows the power released in accreting
    $10^{-6}$~M$_\odot$~yr$^{-1}$ on a Solar-mass protostar of radius
    2~$R_\odot$.  Growing to near Solar mass in a million years
    required the young Sun to reach total luminosities similar to or
    greater than ten times the modern value.
    \label{fig:luminosity}}
\end{figure}

Solar-mass protostars are larger, cooler and more luminous than the
Sun today, according to classical stellar evolution modeling.  The
modeling assumes spherical collapse from some arbitrary large initial
radius, with the mass fixed.  The protostar maintains approximately
constant surface temperature over its first million years, while
shrinking as convection transports the heat of contraction to the
surface where it is radiated away \citep{1961PASJ...13..450H}.  From
``birth'' when the protostar first becomes visible in the infrared,
the luminosity decreases by about an order of magnitude before the
central temperature increases enough for the onset of hydrogen fusion.
The intrinsic luminosity from contraction and nucleosynthesis exceeds
$10 L_\odot$ for the first $2$--$3\times 10^5$~years
(figure~\ref{fig:luminosity}).

The protosun's surroundings were in fact far from spherically
symmetric.  Owing to the difficulty of extracting the angular momentum
from the birth cloud, the material that eventually accreted on the Sun
mostly passed first through the rotation-supported Solar nebula.  Much
of the gravitational energy released in forming the nebula was
radiated away before the gas reached the star.  In the limit where all
the energy escaped, called cold accretion, the protosolar radius was
significantly smaller \citep{1997ApJ...475..770H}.  However if even a
small fraction of the released energy was carried into the protostar
with the accreting material, the protosun was larger and hotter
\citep{2011ApJ...738..140H, 2011arXiv1106.3343H}.

In addition to the intrinsic luminosity, protostars shine with the
energy released during the accretion.  Growing a Solar-mass star in a
million years requires an average accretion rate of $10^{-6}
M_\odot{\rm yr}^{-1}$.  The rate of gravitational energy release near
the end of the growth is $8 L_\odot$ if the accreted material is
deposited on the surface of a star with radius $2 R_\odot$.  If the
accretion is unsteady, the peak luminosities will be higher still.

Empirical stellar temperatures and luminosities are available for some
protostars that have built up much of their mass but are still
accreting gas from their birth cloud.  Near-infrared absorption lines
formed close to the stellar photosphere can be observed despite the
high circumstellar extinction.  An example is YLW~15.  Modeling of its
spectrum indicates the bolometric luminosity $10L_\odot$ is made up of
$3L_\odot$ from the star's internal power sources and $7L_\odot$ from
mass accretion at a rate $1.6\times 10^{-6} M_\odot$~yr$^{-1}$.  The
protostar has mass $0.5M_\odot$ and radius $3.1R_\odot$
\citep{2002AJ....124.2185G}.  Similar results were obtained for larger
samples by \cite{2005A&A...429..543N} and \cite{2005AJ....130.1145D}.
\cite{2009ApJS..181..321E} derived from {\sc Spitzer} photometry an
overall protostellar luminosity function for five nearby star-forming
regions.  The sample in their figure~14 has a median mass well below a
Solar mass, and includes objects aged up to about 0.54~Myr, two to
three times the length of the high-luminosity epoch shown in
figure~\ref{fig:luminosity} above.  Nevertheless a substantial
minority have extinction-corrected bolometric luminosities $10L_\odot$
and above.

Considering the intrinsic and accretion luminosities together, we
conclude that the Sun's total output exceeded $10L_\odot$ for at least
part of the Solar system's first million years.  We investigate the
consequences of the high early Solar luminosity for temperatures in
the surrounding nebula.  Given the uncertainties regarding the
protosolar radius and temperature, it is appropriate to consider
ranges in these parameters.  As representative examples we choose two
model protosuns.  Both are spherical blackbody radiators of mass
$1M_\odot$ and luminosity $10L_\odot$.  Star [A] has radius $5R_\odot$
and temperature $4\,600$~K, while star [B] has radius $2R_\odot$ and
temperature $7\,290$~K.

\subsection{The Solar Nebula
  \label{sec:nebula}}

Measurements using interferometers at sub-millimeter wavelengths show
that protostellar disks have dust surface densities falling off
approximately inversely with radius $R$ outside 30~AU, the smallest
scale resolved in nearby star-forming regions
\citep{2010ApJ...723.1241A}.  The distribution of material nearer the
star has few direct observational constraints.  For the first of three
Solar nebula models, we therefore choose a surface density $\propto
1/R$, normalized so that the outer parts are unstable according to the
Toomre criterion $Q<1$ \citep{1964ApJ...139.1217T,
  1987gady.book.....B}, as required to form Jupiter through a
gravitational instability.  The surface density is
$6\,800$~g~cm$^{-2}$ at 1~AU in this model which we label [D].

As a simple case including a gap, we consider a second model [G] with
an ad hoc reduction in the surface density between 3.3 and 7~AU.  Away
from either gap edge, the surface density declines approximately as a
Gaussian with scale length 5\% of the gap edge radius.  A floor is
also imposed so the surface density is not less than $10^{-6}$ times
that at the same radius in model [D].

For a more detailed model of the disk and gap, we assume viscous
stresses balance the planetary tides, and the planet accretes a
prescribed fraction of the material flowing past
\citep{2006ApJ...641..526L}.  The resulting disk model~[V] has viscous
stress parameter $\alpha=0.005$, planetary accretion efficiency $E=6$,
gap width parameter $f=2$ and planet/star mass ratio $q=1/1000$.  Our
planetary accretion efficiency is lower than the $E=8$ fiducial case
of \cite{2006ApJ...641..526L} because we ultimately obtain a lower
aspect ratio $H/r=0.04$ near 5~AU.

\begin{figure}[tb!]
  \epsscale{0.8} \plotone{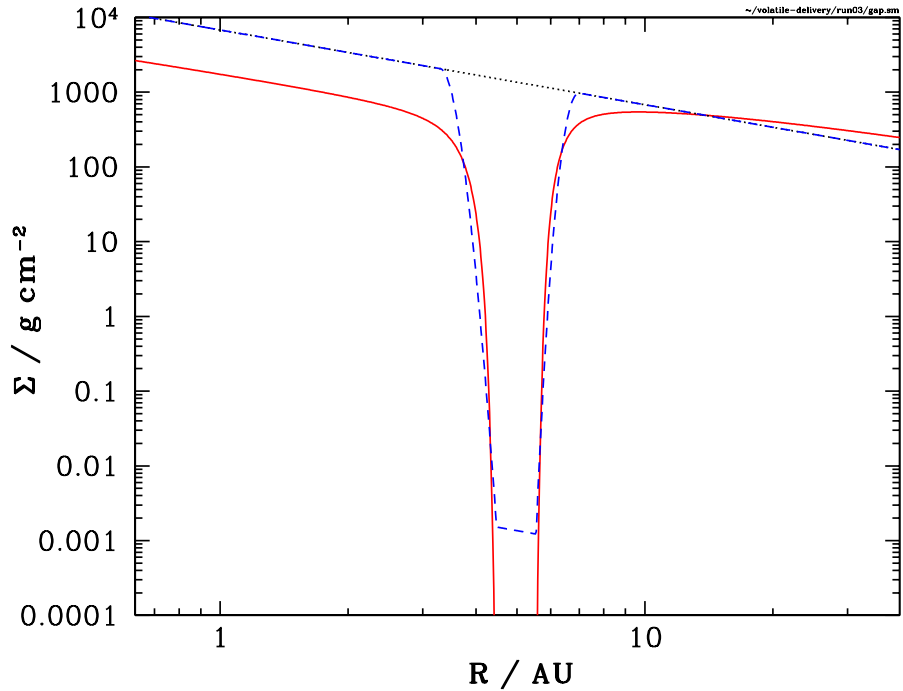} \figcaption{\sf Surface density
    profiles of the three Solar nebula models.  The simple power law
    model [D] is shown by a black dotted line, the power law with gap
    [G] by a blue dashed line, and the viscous disk plus gap model [V]
    by a red solid line.  The viscous model is from Lubow \& D'Angelo
    2005 with planet-to-star mass ratio $q=10^{-3}$, planetary
    accretion efficiency $E=6$ and Shakura-Sunyaev viscosity parameter
    $\alpha=0.005$.  \label{fig:gap}}
\end{figure}

Models [G] and [V] have similar gap widths at unit vertical optical
depth in the starlight (figure~\ref{fig:gap}), with ratios of outer to
inner radius $1.31$.  Comparing the outcomes of the two thus helps
distinguish the effects of the gap from those of the other features
the planet induces in the model [V] surface density profile.

All three nebula models [D], [G] and [V] extend from 0.04 to 40~AU
with a Gaussian roll-off in the surface density at the inner rim to
prevent an artificially hot vertical wall facing the star.  The masses
of the models are 0.191, 0.176 and $0.191 M_\odot$, respectively.  All
three are axially symmetric, and symmetric about the equatorial plane.
The Toomre $Q$ parameter is less than unity indicating gravitational
instability outside 27~AU in disks [D] and [G], and 21~AU in disk [V].

The planet Saturn is also capable of opening a gap in the Solar nebula
under favorable conditions.  Balancing the nebula's viscous stresses
against the tides of a single Saturn-mass planet located at 10~AU,
with all other parameters unchanged, we obtain an alternative version
of model~[V] discussed in section~\ref{sec:discussion}.

In the traditional Shakura-Sunyaev picture, material spiralling inward
through the disk releases its gravitational energy as heat deposited
locally.  This accretional or viscous heating is strong enough to
dominate over the stellar illumination at 5~AU for mass flow rates
above about $10^{-7}$~M$_\odot$~yr$^{-1}$ for T~Tauri star parameters
\citep{2007prpl.conf..555D}.  The higher stellar luminosity we
consider means that accretional heating will be important only at mass
flow rates exceeding $10^{-6}$~M$_\odot$~yr$^{-1}$.  Furthermore,
detailed radiation-MHD calculations of the gravitational energy's
release and dissipation mediated by magnetic fields indicate that most
of the power is deposited at low optical depths in the disk
atmosphere, rather than in the interior \citep{2011ApJ...732L..30H}.
We therefore neglect accretional heating, and obtain lower bounds on
the temperatures by considering only the stellar illumination.

\subsection{Radiative Transfer
  \label{sec:rt}}

We compute the radiative equilibrium temperatures by emitting a large
number of photon packets from the star into the nebula where they are
scattered, absorbed and re-emitted as many times as needed till they
escape to infinity.  With this approach the energy is conserved
exactly.  The stellar luminosity is divided equally among the packets.
We use the temperature relaxation procedure of
\cite{2001ApJ...554..615B}, drawing the frequencies of the re-emitted
packets from the difference between the old and new emission spectra,
such that the local radiation field adjusts to the updated
temperature.  For efficiency when estimating the radiation absorption
rates, we include the contributions from all along the packet paths
\citep{1999A&A...344..282L}.

The nebula's interior is isothermal on cylinders, since we neglect
accretional heating.  We therefore save the expense of computing
temperatures in the most optically-thick regions by simply bouncing
back any packets reaching a certain mass column, chosen so the
overlying material is optically thick at wavelengths near its thermal
emission peak.  The bouncing threshold is set to 20~g~cm$^{-3}$ in all
the calculations shown here.  Additional tests with a threshold of
50~g~cm$^{-3}$ gave similar results.  We replace the missing interior
temperatures by the mean of the last few well-sampled values above.

\begin{figure}[tb!]
  \epsscale{0.8} \plotone{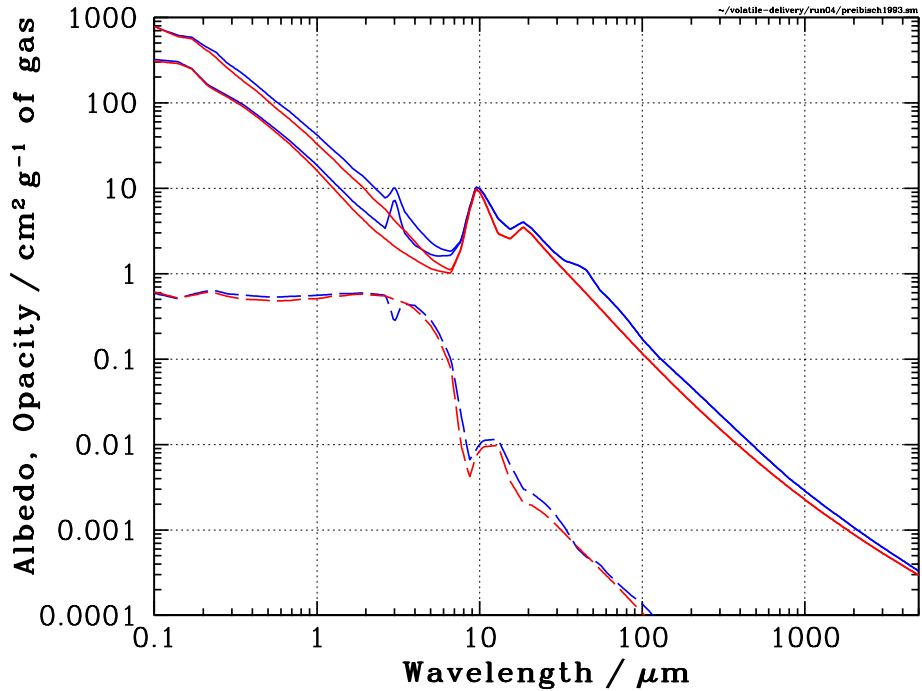} \figcaption{\sf Wavelength
    dependence of the opacities for silicate particles, with (blue)
    and without (red) mantles of ice contaminated with amorphous
    carbon \citep{1993A&A...279..577P}.  The red curves include the
    opacity from the amorphous carbon particles released when the ice
    sublimates.  The lower in each pair of solid curves shows the
    absorption opacity, the upper the sum of absorption and scattering
    opacities.  Dashed curves indicate the albedos.
    \label{fig:opacities}}
\end{figure}

We adopt opacities from \cite{1993A&A...279..577P} who matched Mie
calculations of dust particles' optical response against data from
molecular clouds.  The grain model consists of silicate particles,
mantled with water and ammonia ice that is polluted with fine
carbon-rich particles.  The mantles increase the particles' radii by
14.5\%.  At temperatures above 125~K the ice sublimates, releasing the
carbon particles.  Above $1\,500$~K the silicates also sublimate,
leaving only the carbon particles.  Finally, above $2\,000$~K the
carbon particles are destroyed too.  Each particle species has a
power-law size distribution with exponent $-3.5$.  The minimum and
maximum sizes are 0.04 and 1~$\mu$m for the silicate particles and
0.007 and 0.03~$\mu$m for the carbon particles.  The opacity curves
are shown in figure~\ref{fig:opacities}.  Scattering contributes about
half of the total cross-section at optical wavelengths, and is assumed
isotropic.  Above $2\,000$~K we set the opacity to a low level of
$10^{-4}$~g~cm$^{-3}$ representing the residual effects of molecular
lines in the opacity gap at temperatures between the dust destruction
and hydrogen ionization thresholds \citep{1994ApJ...427..987B}.  Such
high temperatures are found only far from the gap, in the
optically-thin material near the disk's inner rim.

The disk is divided into a grid of cells, each of which has a single
uniform temperature for dust and gas.  The dust is assumed to be
well-mixed in the gas.  The grid has 300~cells spaced logarithmically
in radius $R$ between 0.04 and 40~AU.  With 100~cells per decade, each
cell is 2.3\% further from the star than the last.  In the vertical
direction, the grid has 200~cells uniformly-spaced between the
equatorial plane and height $z=0.4 R$, yielding cells separated by
0.2\% of the radius.

\subsection{Joint Radiative and Hydrostatic Equilibrium
  \label{sec:hse}}

Once new temperatures have been found through the Monte Carlo
radiative transfer procedure, we restore vertical hydrostatic
equilibrium by solving the force balance equation within each disk
annulus, while holding fixed the surface density and the variation of
the temperature with the column mass.

We iterate five times between radiative transfer and hydrostatic
balancing.  Each iteration involves $10^6$~photon packets.  We then
use the result of the fourth iteration as the initial condition for
two further iterations with $10^7$~packets each, and check that the
larger number of packets leaves the disk structure almost unchanged.

During the thermal relaxation procedure the opacity curves must remain
fixed \citep{2001ApJ...554..615B}, while the opacities we use vary
with temperature across the sublimation thresholds for water,
silicates and carbon-rich grains (figure~\ref{fig:opacities}).
Throughout each iteration we therefore set each grid cell's opacity
curve according to the temperature found there on the previous
iteration.

\section{RESULTS
  \label{sec:results}}

\begin{figure}[tb!]
  \epsscale{0.8} \plotone{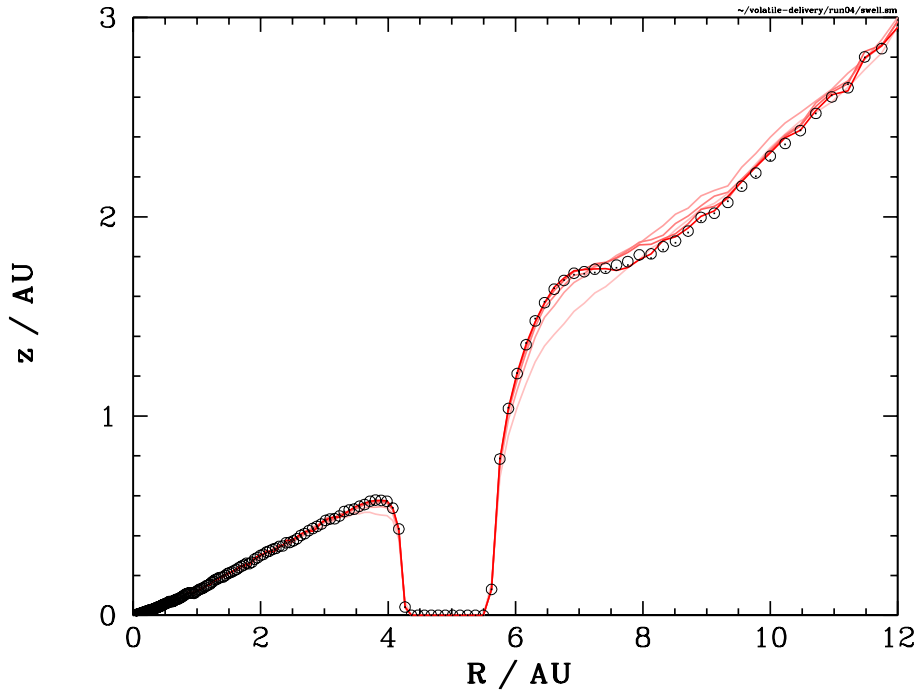} \figcaption{\sf Surface of unit
    vertical optical depth at the wavelength of the stellar blackbody
    peak (0.65~$\mu$m) after one, two, three, four and five iterations
    of radiative and hydrostatic balancing (curves colored from pink
    to deep red) in run~3.  Each iteration involves $10^6$~photon
    packets.  The height of the gap's outer edge near 6.5~AU increases
    over the first three iterations and then changes little between
    the last three.  We also use the fourth iteration as the initial
    condition for two further iterations with $10^7$~packets each,
    shown by black circles and black points.  The increased number of
    packets leaves the disk structure almost unchanged.
  \label{fig:swell}}
\end{figure}

\begin{figure}[tb!]
  \epsscale{0.8} \plotone{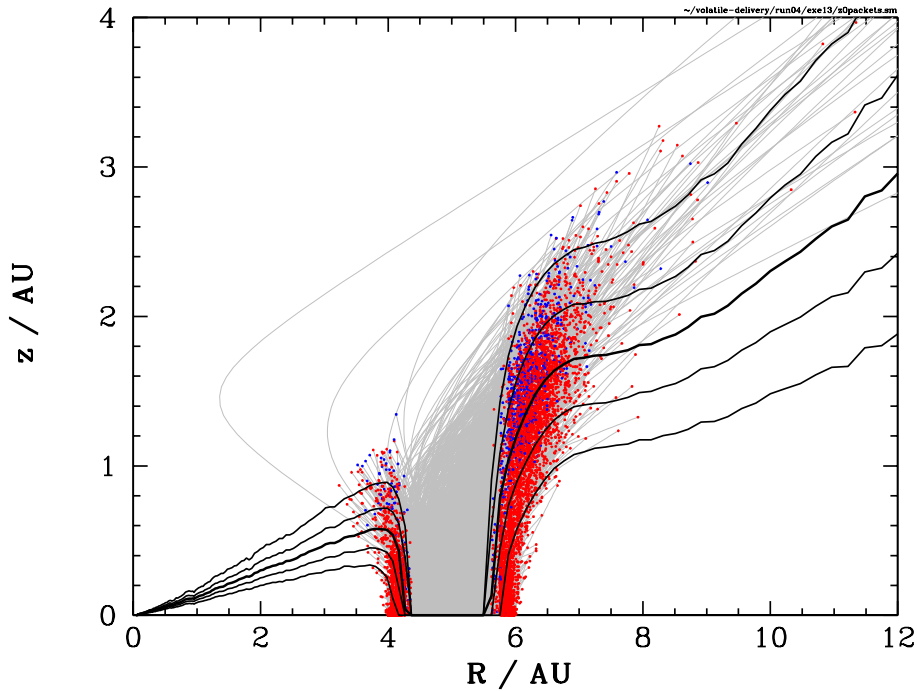} \figcaption{\sf Paths of every tenth
    photon packet reaching the gap midplane in run~3.  The number of
    packets plotted is~7613.  Dots mark the locations of last
    interaction before crossing the midplane, and their colors show
    whether the packet was last absorbed (red) or scattered (blue).
    All packets carry equal energy.  Black lines indicate the surfaces
    of vertical optical depth 0.01, 0.1, 1 (heavier line), 10 and 100
    (top to bottom) at the stellar blackbody peak wavelength
    (0.65~$\mu$m).
    \label{fig:z0packets}}
\end{figure}

\begin{figure}[tb!]
  \epsscale{0.8} \plotone{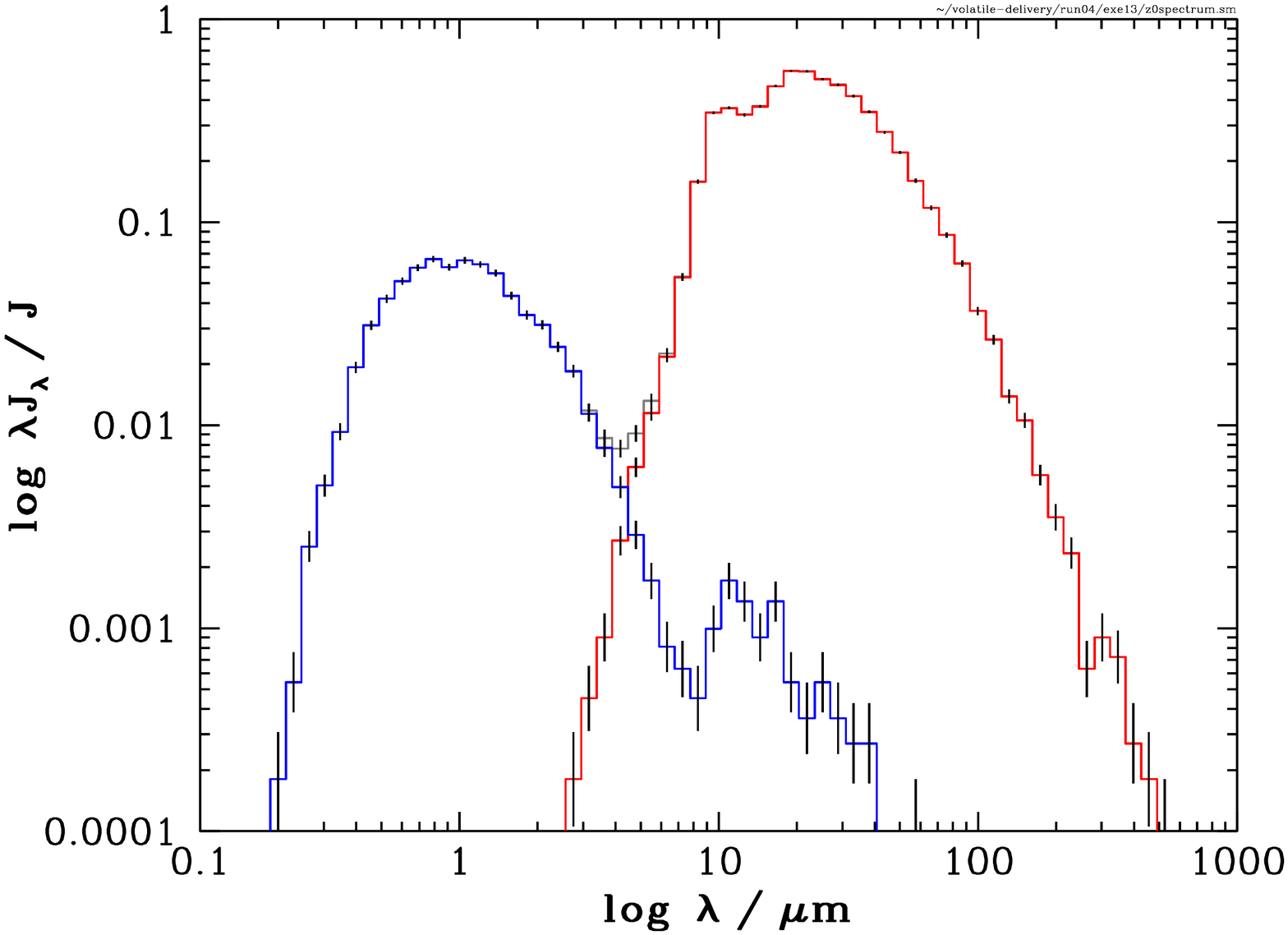} \figcaption{\sf Spectrum of the
    radiation reaching the gap midplane in the same calculation shown
    in figure~\ref{fig:z0packets}.  The packets that most recently
    interacted by scattering (blue histogram) have a spectrum similar
    to the star, while the packets that were last absorbed and
    re-emitted (red) are at mid- to far-infrared wavelengths
    corresponding to the temperatures found in the disk.  A weak
    silicate feature appears at 10-$\mu$m.  The grey histogram visible
    only at intermediate wavelengths is the sum of the scattered and
    absorbed radiation.  The vertical axis is the mean specific
    intensity $J_\lambda$ as a fraction of the mean bolometric
    intensity $J$, multiplied by the wavelength so that equal areas
    correspond to equal luminosities.  The resulting quantity is
    dimensionless.  Error bars show the $\sqrt{N}$ uncertainties under
    Poisson statistics.
    \label{fig:z0spectrum}}
\end{figure}

\begin{figure}[tb!]
  \epsscale{0.8} \plotone{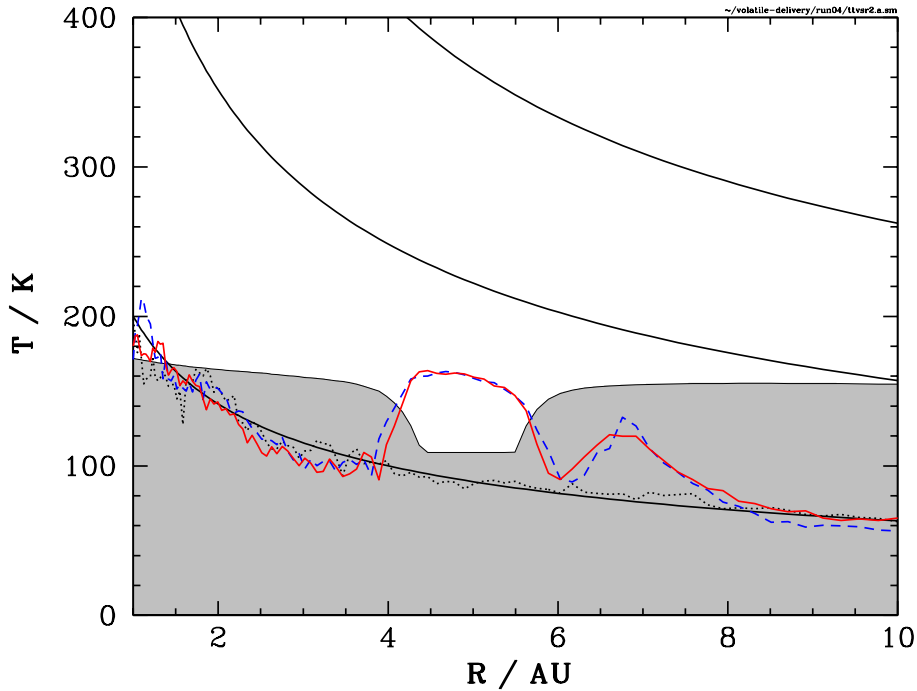} \figcaption{\sf Midplane temperature
    vs.\ radius in the runs with star~[A].  Dotted, dashed and solid
    lines show respectively runs 1, 2 and~3.  Grey shading marks the
    region where water ice is stable at the midplane densities found
    in run~3.  Solid black curves show for comparison the temperatures
    of (top to bottom) grains directly exposed to the starlight,
    blackbodies directly exposed to the starlight, and the initial
    condition for the radiative-hydrostatic iterations, $T=200\,{\rm
      K}/R_{\rm AU}^{1/2}$.  \label{fig:ttvsra}}
\end{figure}

\begin{figure}[tb!]
  \epsscale{0.8} \plotone{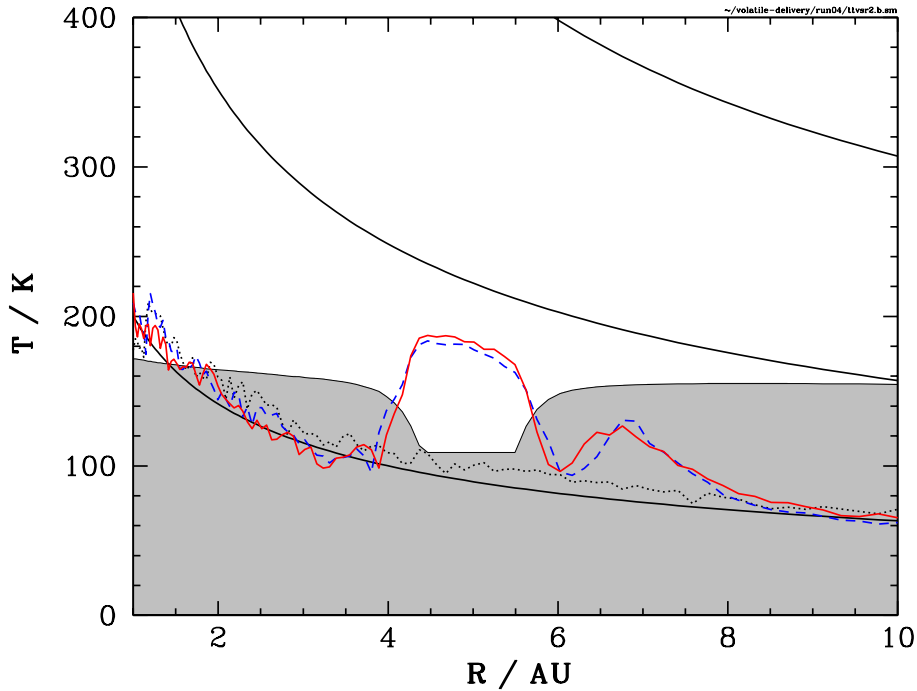} \figcaption{\sf Same as
    figure~\ref{fig:ttvsra} but for the runs 4, 5 and 6 with star~[B].
    The shorter wavelengths of the light emitted by the hotter star
    correspond to higher dust opacities, leading to a hotter gap than
    in figure~\ref{fig:ttvsra}.  Temperatures are higher also for
    grains directly exposed to starlight (solid black curve at top
    right).  \label{fig:ttvsrb}}
\end{figure}

\begin{figure}[tb!]
  \epsscale{0.8} \plotone{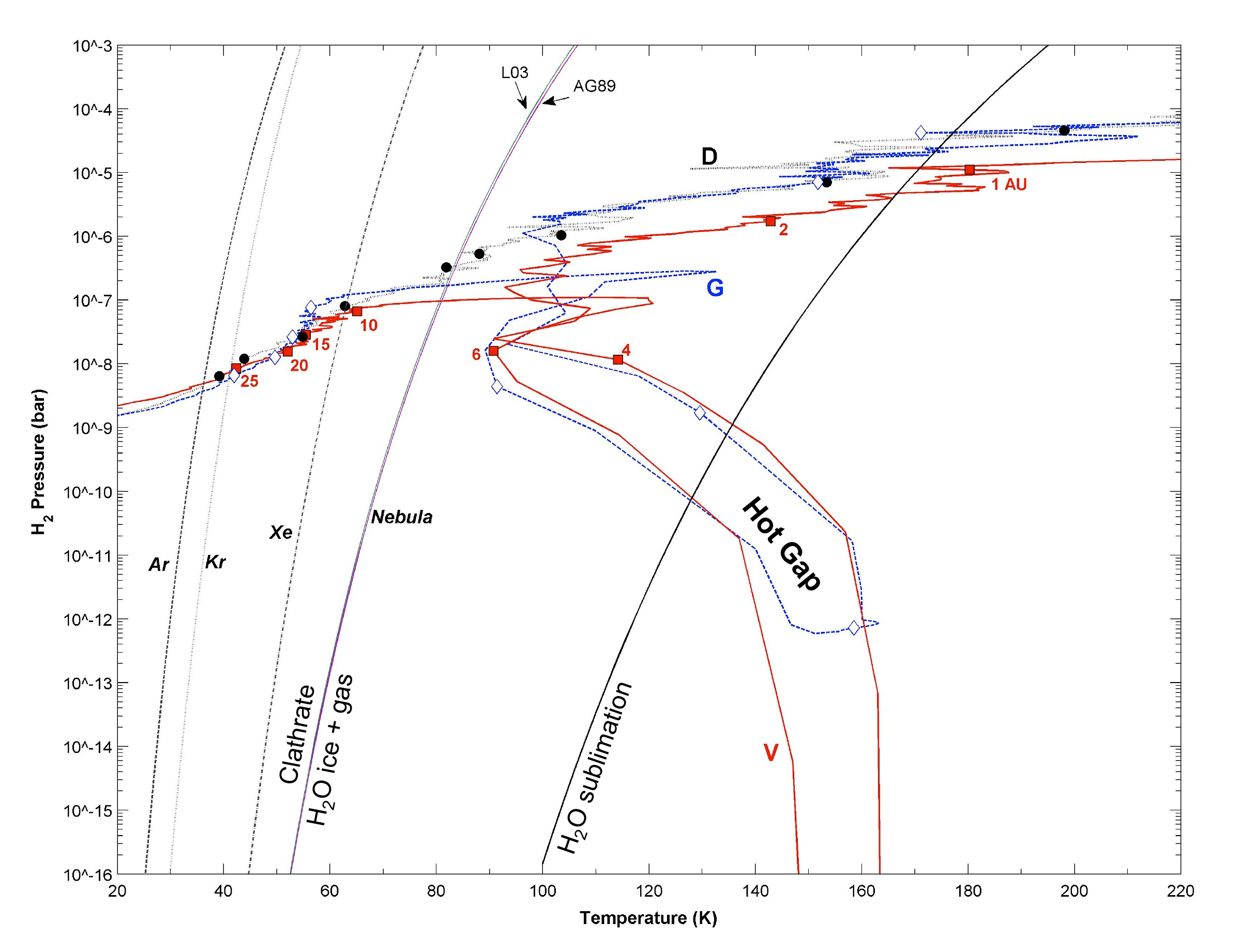} \figcaption{\sf Pressure-temperature
    profiles of the three nebula models with star~[A], compared
    against the sublimation curve of water ice and the stability
    curves of clathrate hydrates.  As in previous figures, dotted,
    dashed and solid lines are for the models with disks [D], [G] and
    [V], respectively.  Heliocentric distance markers are placed at 1,
    2, 4, 5, 6, 10, 15, 20, and 25~AU.  The curves are shown for pure
    argon, krypton and xenon clathrates, and for a mixed clathrate in
    equilibrium with the solar nebula for elemental abundances after
    \cite{1989GeCoA..53..197A} and \cite{2003ApJ...591.1220L}, labeled
    AG89 and L03 respectively.  In all three models, clathrates are
    unstable at Jupiter and thus cannot readily deliver volatiles to
    the planet.  Water ice sublimates easily in the two models with a
    gap.
\label{fig:ptplane}}
\end{figure}

\begin{figure}[tb!]
  \epsscale{0.8} \plotone{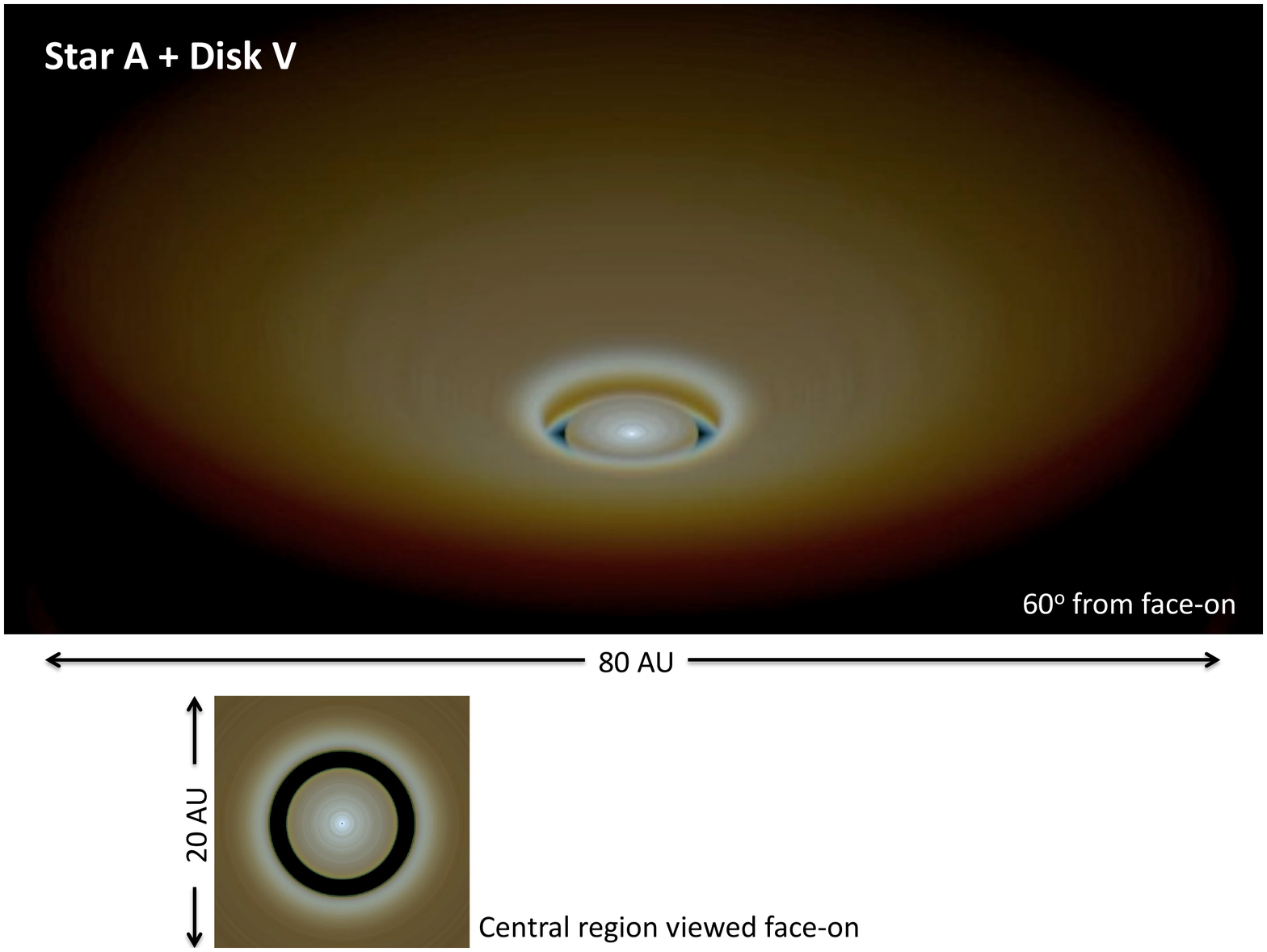} \figcaption{\sf Synthetic images of
    run~3.  The red, green and blue channels correspond to wavelengths
    40, 20 and 10~$\mu$m.  The intensity scale is logarithmic with the
    central protostar saturated.  At top, the system is viewed
    $60^\circ$ from face-on and the field of view is just over 80~AU
    across.  The gap's outer rim is hotter than its surroundings and
    appears as a bright ring.  Also visible through the gap is the hot
    rim on the disk's lower surface.  The gap itself appears bluish
    because the sightlines pass through optically-thin hot material
    overlying the rings that emits most strongly at 10~$\mu$m.  The
    bottom panel is a face-on view of the central 20~AU on the same
    color scale.
    \label{fig:images}}
\end{figure}

We compute temperatures in the gap for all six star-disk combinations.
The six runs are listed in table~\ref{tab:runs} along with the
temperatures $T_p$ found at the planet's orbit.  The temperatures in
the equatorial plane are especially important for the ice
distribution, since the vertical component of the Sun's gravity keeps
macroscopic solid bodies lying mostly near the nebula's midplane
\citep{1993prpl.conf.1031W}.  All four cases with gaps have similar
temperature profiles, as do the two cases without gaps.  We choose to
focus on run~3 with the larger, cooler star~[A] and viscous disk~[V].
Selected results are in figures~\ref{fig:swell} to~\ref{fig:images}.

\begin{table}[tb]
  \caption{\sf The six Monte Carlo radiative transfer calculations.
    \vspace*{3mm}
    \label{tab:runs}}
\begin{center}
\begin{tabular}{lllr}
\hline
Run   & Star  & Disk  & $T_p$ \\
number& model & model & (K)   \\
\hline
\hline
1         & [A]  & [D]  &  88 \\
2         & [A]  & [G]  & 159 \\
3         & [A]  & [V]  & 159 \\
\hline
4         & [B]  & [D]  & 100 \\
5         & [B]  & [G]  & 178 \\
6         & [B]  & [V]  & 183 \\
\hline
\end{tabular}  
\end{center}
\end{table}  

Several iterations of the radiative and hydrostatic equilibrium
procedure for run~3 are shown in figure~\ref{fig:swell}.  The
temperature is initially set to $200\,{\rm K}/R_{\rm AU}^{-1/2}$,
approximating the final midplane temperature profile obtained in
run~1.  The disk is thus initially isothermal on cylinders.  Its
flared surface together with the absence of material from the gap
means the gap's outer edge receives extra starlight.  Note that the
effect is exaggerated in figure~\ref{fig:swell} by the magnified
vertical scale.  The gap's outer edge heats and, under hydrostatic
equilibrium, expands.  On the next iteration, the taller outer edge
intercepts yet more starlight and expands further.  At the same time,
the gap's inner edge receives reprocessed radiation from the outer
edge and also heats and expands slightly.  The changes are smaller
with successive iterations, and after three iterations the gap changes
little.  However the disk beyond the gap's puffed-up outer edge now
lies in shadow, and becomes cooler and thinner.  After falling into
shadow, the outer disk also rapidly converges, and between the two
iterations with $10^7$~packets, there is little change.

As described in section~\ref{sec:hse} our procedure yields two
versions of the fifth iteration, one using $10^6$~photon packets and
another with $10^7$.  The differences between these two are consistent
with the expected Monte Carlo noise in all six runs.  Therefore from
now on we discuss only the results from the version of the fifth
iteration with $10^7$~packets.

The paths followed by packets entering the gap are plotted in
figure~\ref{fig:z0packets}.  Most of the radiation heating the gap
midplane is starlight reprocessed in the outer rim, as shown by the
larger number of red than blue points.  The packet paths appear curved
because for purposes of the plot, they are projected onto a meridional
plane by rotation.  Some of the packets enter the gap traveling near a
tangent to the planet's orbit.

The spectrum of the radiation field in the gap is in
figure~\ref{fig:z0spectrum}.  While most of the radiation lies in the
peak corresponding to the disk's thermal emission, a significant
minority (11\% of the power) is in scattered starlight.  A further
fraction of a percent is absorbed and re-emitted by the disk, and
subsequently scattered before reaching the gap (blue curve, right-hand
end).  The bolometric mean intensity of the radiation at the planet's
orbit is 3.7\% of that at the same location if the intervening disk
material were removed.

The midplane temperature profiles from the six runs are plotted in
figures~\ref{fig:ttvsra} (for the three runs with star~[A])
and~\ref{fig:ttvsrb} (star~[B]).  Despite the different surface density
profiles of models [G] and [V], runs 2 and 3 have quite similar
temperatures in the gap.  The same is true of runs 5 and 6, where the
temperatures are generally higher, due to the greater opacity of the
dust at the shorter wavelengths emitted by the hotter star~[B].

The grey shading in figures~\ref{fig:ttvsra} and~\ref{fig:ttvsrb}
indicates the temperature range where water ice is stable at the
midplane density found in the run with the viscous disk [V].  The gaps
reach temperatures well above the water sublimation thresholds in all
four runs with gaps, as shown by the curves exiting the gray shaded
regions.  We took the dependence of the sublimation temperature vs.\
density from \cite{1994ApJ...421..615P}, choosing the case where water
ice contains 52\% of the available oxygen.  Lower fractions of the
oxygen locked up in the water would correspond to lower sublimation
temperatures.  In the gap center in model [V] the densities are less
than the lowest value in the Pollack et al. table,
$10^{-18}$~g~cm$^{-3}$.  Although the true sublimation temperature
should be lower, we simply use the lowest tabulated value.

Heavy elements can be delivered to the planets trapped in ices,
provided the material is kept at temperatures somewhat below the ice
sublimation threshold \citep{1985ApJS...58..493L}.  In order to
determine the heliocentric distance where trapping is effective, we
calculate the stability curve of a mixed clathrate hydrate in
equilibrium with the solar nebula, for two commonly-used
determinations of the nebular elemental abundances
\citep{1989GeCoA..53..197A, 2003ApJ...591.1220L}.

First, we apply empirical expressions fitted to experimental data from
the literature to calculate the dissociation pressure $P_{d,i}$ as a
function of temperature for the $i$ species CH$_4$, CO$_2$, Xe
\citep[parameters from][]{2010JCED...55.5101F}, CO, N$_2$, H$_2$S, Ar,
and Kr \citep[parameters from][]{2004P&SS...52..623H}.  Then we use
the method described in \cite{2009ApJ...696.1348M} to determine the
molar fractions $y_i$ of each species in the gas phase, for elemental
abundances taken from \cite{1989GeCoA..53..197A} and
\cite{2003ApJ...591.1220L}.  The stability curve of a mixed clathrate
in equilibrium with the solar nebula is obtained by calculating the
dissociation pressure at each temperature, $P_{d,mix}$ using the
following equation \citep[e.g.][]{1974ngms.book..151M}:
\begin{equation}
P_{d,mix} = \left(\sum_i{y_i\over P_{d,i}}\right)^{-1}
\end{equation}
The resulting stability curves are plotted in figure~\ref{fig:ptplane}
along with the temperature-pressure profiles of three of our model
solar nebulae.  The stability curves for the two sets of elemental
abundances are very similar, and both indicate clathrate hydrates are
stable in the three nebulae at temperatures below 81~K found only
outside 7.7~AU, well beyond the outer edge of the gap.  Furthermore,
these stability curves are extremely close to the stability curve of
pure H$_2$S clathrates (not shown for legibility), which implies that
clathrate hydrates forming under these conditions contain almost
exclusively H$_2$S.  Much colder temperatures, below about 50~K, must
be attained for these clathrates to trap large amounts of Ar, Kr, Xe
and N$_2$.  Therefore, the volatile fraction of planetesimals that
formed Jupiter needs to originate from heliocentric distances
exceeding 20~AU to account for the heavy element enrichments measured
by the Galileo spacecraft in the planet's atmosphere
\citep[e.g.][]{1999Natur.402..269O}.

Finally, synthetic images of the run~3 model nebula, of the kind that
might one day be obtained by pointing a large infrared interferometer
at a protostellar disk in a nearby star-forming region, are shown in
figure~\ref{fig:images}.  The gap's outer edge is hot because it is
both directly exposed to starlight, and more tilted toward the star
than other parts of the disk.  Furthermore, it is surrounded on the
inside by the optically-thin dark gap and on the outside by disk
annuli in shadow.  As a result the gap's outer edge stands out as a
bright ring.  At wavelength 20~$\mu$m, near the peak of the local
thermal emission, the face-on surface brightness contrast between the
bright ring at 6.5~AU and the shadowed disk at 10~AU is a factor 23.

\section{DISCUSSION\label{sec:discussion}}

Below we discuss the robustness of the hot gap phenomenon, the
distribution of the heating, the consequences for primitive bodies in
and near the gap, and several scenarios for the volatile enrichment of
Jupiter and its satellites.

\subsection{Location and Extent of the Heating\label{sec:heatingextent}}

The high Solar luminosity makes the whole nebula hotter than a T~Tauri
disk, even in the absence of any gaps.  The temperature at Jupiter's
orbit is high enough to destroy clathrate hydrates, at 88 to 100~K
(table~\ref{tab:runs}; figure~\ref{fig:ptplane}).  There will be
consequences also for the stability of amorphous ice, whose presence
near the giant planets has already been questioned
\citep{1994A&A...290.1009K, 2005SSRv..116...25G}.  With the potential
carrier materials unstable, it will be difficult to form the planet in
situ from material enriched in volatile-trapping ices.

Stars less luminous than $10L_\odot$ yield lower temperatures, but the
gap remains above the water sublimation threshold even at $3L_\odot$,
as we check using the following scaling.  The flux of radiation
heating the gap is proportional to the Solar luminosity $L$ and to the
disk thickness.  The thickness in turn is proportional to the sound
speed, which varies as the square root of the temperature.  On the
other hand, the radiative cooling rate of the material in the gap is
proportional to $T^4$ times the Planck-mean absorption opacity
$\kappa_P$.  At temperatures near the sublimation threshold,
$\kappa_P\sim T^{5/4}$ approximately.  Balancing heating and cooling
yields $T\sim L^{4/19}$, so reducing the luminosity from 10 to
$3L_\odot$ should reduce the temperature from 159 to 123~K.  This is
close to the temperature of 124~K found at the planet's orbit, in a
version of run~3 with the luminosity reduced to $3L_\odot$ by
shrinking the stellar radius to $2.75R_\odot$.  With the luminosity
reduced further to $1L_\odot$ by shrinking the star to $1.59R_\odot$,
the temperature of 101~K again is close to the scaling law estimate of
98~K.

Our temperatures are lower limits since we neglected both the
luminosity of the protoplanet and the accretion luminosity of the
nebula itself.  We also neglected the contribution to the radiation
field from the envelope of gas and dust surrounding and falling onto
the nebula.  Absorption by material along the line of sight from the
star can reduce the flux reaching the disk, while reprocessing by
material above the disk increases the flux
\citep{1997ApJ...474..397D}.  Estimating the net effect is a
worthwhile goal for future work.  We have neglected the anisotropy of
the scattering, but that is unlikely to change the temperatures
significantly because most of the photon packets reaching the planet's
orbit were absorbed and re-emitted rather than scattered
(figure~\ref{fig:z0spectrum}).

The temperatures inside the gap are basically independent of the
surface density, provided the gap is optically-thin in the vertical
direction, as shown by the close match between the results with
disks~[G] and [V] (figures~\ref{fig:ttvsra} and~\ref{fig:ttvsrb}).
The outcome is thus independent of the density floor chosen in
model~[G].  In the same vein, optically-thin non-axisymmetric wakes
raised within the gap by the planet's tides are likely to have little
effect on the heating.  Unit optical depth corresponds to surface
density a few tenths of a gram per square centimeter, considering the
opacity (figure~\ref{fig:opacities}) at the radiation field's peak
wavelength of 20~$\mu$m (figure~\ref{fig:z0spectrum}).  The
temperatures also increase only slowly with height, being less than
3~degrees hotter at 0.05~AU above the planet.  On the other hand, the
outcome will depend on the gap width.  The outer rim of a narrower gap
will receive less direct sunlight, and re-radiate less power toward
the equatorial plane.

The gap is optically-thin to the sunlight in the vertical direction
between 4.33 and 5.68~AU in our run~3, while the temperatures computed
by the Monte Carlo procedure exceed the local sublimation threshold
from 4.17 to 5.62~AU.  Note that the high temperatures extend past the
edge of the optically-thin region only on the gap's inner side, which
more directly views the sunlit outer rim.  The heated region includes
the orbits of Jupiter's Trojan asteroids, the Thule group and the
outer parts of the Hilda group.  Any primordial populations in these
locations will be thermally altered.  It would be interesting to learn
whether the planetesimal devolatilization associated with an early
origin of Jupiter is compatible with the present makeup of the
asteroid families, under standard scenarios for Solar system dynamical
evolution such as the Nice model \citep{2005Natur.435..459T}.


Bodies trapped beyond the gap's outer rim also experience ambient
temperatures high enough to affect volatile species.  Temperatures
near the local maximum in gas pressure, at 6.8~AU, are 120 to 133~K in
the four models from table~\ref{tab:runs} with gaps, owing to the
extra sunlight absorbed on the rim overhead.  Gas drag forces push
meter-sized bodies toward the pressure maximum
\citep{2000ApJ...540.1091B, 2003ApJ...583..996H}, where they can spend
many orbits in warm surroundings.  The temperatures and pressures are
in the regime where water ice is stable but clathrate hydrates are not
(figure~\ref{fig:ptplane}).  At the same location with no gap, the
temperature is about 83~K and some of the clathrates are stable.  In
general there is a band at or beyond the planet's orbit where H$_2$S
and xenon clathrates are stable, but argon clathrates are not.  This
is problematic since Jupiter is enriched in many volatile elements
including sulfur, xenon and argon by similar factors of about~2.5 with
respect to Solar \citep{2003NewAR..47....1Y}.

Jupiter could have formed early but further from the star, where the
temperatures were low enough for water ice and clathrates to be
stable.  In this scenario, the planet later migrated inward to its
present position.  We investigate the migration distance required
using a pair of additional experiments.  Star~[A] illuminates a
version of the viscous disk~[V] with the planet moved to 10~AU in one
case, and 20~AU in the other.  The temperatures reached in the gap at
the planet's orbit are 117 and 84~K, respectively.  The first is above
the water sublimation threshold, while the second allows stable water
ice but destroys clathrate hydrates.  Enriching Jupiter in heavy
elements that were trapped in clathrates in and near the planet's gap
was difficult during the early epoch we consider, unless the planet
formed outside 20~AU.

Saturn too can be located in a hot gap, if its tides open a clearing
in the nebula that is wide enough for a substantial area on the outer
rim to be directly sunlit, and if the optical depth is sufficiently
low for the reprocessed and scattered radiation to reach the planet.
We evaluate the temperatures using a further run having star~[A] and a
version of disk~[V] in which a single Saturn-mass planet orbits at
10~AU.  Due to its lower mass, Saturn opens a gap almost three times
narrower than run~3 in relative terms, with the outer radius at unit
vertical optical depth only 11\% greater than the inner.  The gap also
has weaker surface density contrast and, although still quite
optically thin to its own thermal emission, is only marginally
optically thin for photons at the stellar blackbody peak, with an
optical depth of 0.18 at the planet's orbit.  The resulting
temperature is 100~K, well below the value of 117~K obtained with
Jupiter at the same location.  Water ice is then stable, but
clathrates decompose.

\subsection{Enriching Jupiter in Volatiles}

Jupiter can potentially form directly from material enriched in heavy
elements in a gravitationally unstable Solar nebula.  The nebula
develops spiral arms where gas drag forces concentrate meter-sized
solid bodies \citep{2004MNRAS.355..543R}.  The planets form in part
from the arms and so receive extra boulder-sized material compared to
the nearby nebula \citep{2010ApJ...724..618B}.  However we have seen
that temperatures even in the optically-thick early nebula are high
enough to make volatile-trapping ices unstable
(section~\ref{sec:heatingextent}).  It will be difficult to form
Jupiter at 5~AU directly from material enriched in volatile elements
such as the noble gases.

As the planet nears its final mass, its tidal forces become capable of
opening a hot gap.  However, bringing the tides into balance with the
nebula's accretion stresses takes a few $10^4$-year accretion
timescales across the gap width, much longer than the $10^3$-year
timescale for growing the planet by gravitational collapse
\citep{1997Sci...276.1836B}.  The gap will therefore clear only after
Jupiter receives nearly all its mass.  If the planet is not yet
enriched on reaching this stage, newly added material must be more and
more depleted in hydrogen and helium in order to produce the final
abundances.  Once it becomes insufficient to accrete nebular gas mixed
with the vapors from sublimated volatile ices, the last viable route
to enrichment is by accreting the solid icy bodies themselves.

The impact of the hot gap on icy bodies in Jupiter's vicinity depends
on the bodies' sizes:
\begin{enumerate}
\item Bodies small enough to sublimate within the minimum gap-crossing
  time (about one orbital period or 10~yr) lose their clathrates,
  gas-laden amorphous ice, and other volatiles, and cannot deliver
  noble gases.  The largest solid icy body that sublimates completely
  on this timescale at temperature 160~K is about 10~cm, using the
  sublimation law from \cite{1993PhRvB..48.9973S}.  Smaller bodies
  lose their clathrates, gas-laden amorphous ice, and other volatiles,
  and cannot deliver noble gases.
\item Somewhat bigger icy bodies heat to the ambient temperature
  within the minimum gap-crossing time but do not completely
  sublimate.  The smallest body capable of keeping a cool center
  during the gap-crossing is 10~m, using 0.01~cm$^2$~s$^{-1}$ for the
  thermal diffusivity of water ice.
\item Larger planetesimals could preserve volatiles.  Bigger bodies'
  internal temperatures remain below the clathrate loss threshold for
  at least an orbital period when exposed to the gap.  These bodies
  can deliver noble gases to Jupiter if they reach the planet fast
  enough.  However, icy planetesimals that remain in the gap for
  extended time periods eventually vaporize unless their ice is
  shielded by the development of a lag deposit.  This can be achieved
  if the regolith reaches a depth of a few hundred meters
  \citep{1989Icar...82...97F, 2008ApJ...682..697S}.
\end{enumerate}
Dynamics impose an additional constraint on the sizes of the bodies
reaching the planet.  Bodies larger than a meter can be trapped in
outer mean-motion resonances with the planet
\citep{1985Icar...62...16W}, while particles of centimeter to meter
size are pushed away from the planet's orbit by gas drag forces.  The
bodies most easily accreted are grains smaller than 10~$\mu$m, which
are swept in with accreting gas \citep{2007A&A...462..355P}.  If
applied together, the dynamical and thermodynamical constraints
prevent all enrichment in volatiles with respect to the nearby Solar
nebula: only small particles approach the planet, and these quickly
vaporize.

The hot gap also illuminates the circumjovian nebula, raising
temperatures to levels that may impact the chemistry of the forming
Jovian satellites.  Growing Ganymede requires conditions appropriate
for the condensation of water ice \citep{2010tfch.book...35L}, well
below the temperature floor imposed by the radiation bathing the
planet's environment in the hot gap.  The circumplanetary material
will reach temperatures similar to those listed in
table~\ref{tab:runs} whether optically-thick
\citep{2003Icar..163..198M} or thin \citep{2002AJ....124.3404C}.  Our
findings thus indicate that an early satellite formation epoch is
inconsistent with the Ganymede temperature constraint.  In any case,
it has been suggested that the Galilean satellites correspond to a
late generation of moons formed around the time the Solar nebula
dispersed \citep{2006Natur.441..834C, 2008Icar..198..163B}, several
million years after the hot gap event.

\section{Conclusions\label{sec:conclusions}}

We used a Monte Carlo radiative transfer approach to compute
temperatures in the Solar nebula during the early epoch when the Sun
was ten times more luminous than today.  The midplane around 5~AU was
hot enough to destroy clathrate hydrates, even with the nebula
optically-thick, pointing to difficulties in forming Jupiter in situ
directly from material enriched in volatiles such as the noble gases.

If the planet nevertheless approached its final mass within a few
hundred thousand years of the Sun, opening an optically-thin gap in
the Solar nebula, then enough sunlight was reprocessed and scattered
into the gap to heat the gas and dust to temperatures above 150~K that
caused the loss of water and other volatiles from incoming small
bodies, slowing or stopping the enrichment of the planet in heavy
elements with respect to its surroundings.

An early origin may still be possible, if Jupiter either (1) formed
much further from the Sun, and was enriched before migrating inward,
or (2) was enriched in heavy elements and acquired its icy satellites
only later, after the Sun became less luminous.  In the first case,
Jupiter would need to form beyond 20~AU in the particular model we
considered, to enable the delivery of clathrate hydrates bearing the
noble gases.  In the second case, the enrichment must occur after the
Sun has dimmed below $3L_\odot$, by which time the planet's cooling
and contraction can greatly reduce its cross-section for capturing
planetesimals \citep{2011Icar..211..939H}.

More generally, our results demonstrate that the Sun's luminosity is a
parameter worth accounting for in any model of the formation of
volatile-rich bodies.

\acknowledgements

We thank T.~Hosokawa and H.~Yorke for discussions regarding
protostellar evolution.  The research was carried out at the Jet
Propulsion Laboratory, California Institute of Technology, under a
contract with the National Aeronautics and Space Administration.  The
project was supported by the JPL Research \& Technology Development
program, and by the NASA Outer Planets Research program under grant
07-OPR07-0065 to N.~J.~T.  Copyright 2011 California Institute of
Technology.  Government sponsorship acknowledged.

\bibliographystyle{apj}
\bibliography{ysodisk}

\end{document}